# An Educative Brain-Computer Interface

Kirill Sorudeykin

*Kharkov National University of Radio Electronics*

*Kirill.A.Sorudeykin@ieee.org*

*Physical concepts are free creations of the human mind, and are not, however it may seem, uniquely determined by the external world.*

ALBERT EINSTEIN, *The Evolution of Physics*

## Abstract

*In this paper we will describe all necessary parts of Brain-Computer Interface (BCI), such as source of signals, hardware, software, analysis, architectures of complete system. We also will go along various applications of BCI, view some subject fields and their specifics. After preface we will consider the main point of this work – concepts of using BCI in education. Represented direction of BCI development has not been reported prior. In this work a computer system, currently being elaborated in author's laboratory, will be specified. A purpose of it is to determine a degree of clearness of studied information for certain user according to their indications of brain electrical signals. On the basis of this information the system is able to find an optimal approach to interact with each single user. Feedback individualization leads to learning effectiveness increasing. Stated investigations will be supplemented by author's analytical reasoning on the nature of thinking process.*

## 1. Introduction

In 1924 Hans Berger, a German psychologist, have made the first human electroencephalography (EEG) recording. Since that time EEG became wide popular as an instrument for medical diagnostics and scientific investigations. It provides a possibility of non-invasively measure an electrical activity of regions of brain cells from the surface of a scalp.

Electroencephalograph has analog and digital circuits. Similarly to lots of medical devices, it contains such elements as a protection circuit, an operational amplifier, several high frequency and low frequency filters to remove various kinds of noise, ESD-protection circuits, a microcontroller and some other computer chips. Device should be electrically safe and must correspond to the modern standards such as IEC 60601-1. Various types of EEG-hardware can be designed in the form of embedded system or single device, for example, neuroprosthetics or lightweight physiological data recorders. Electrical potentials are acquired by electrodes, connected to the EEG device. Group of electrodes gather in a special electrode cap. The location of scalp electrodes is described in corresponding standards. Electrode basis is usually made of gold, silver, silver-chloride, steel, etc. Electrodes can be active or passive. Active electrodes contain built-in amplifier to reinforce biopotentials directly on the place of acquisition.

Control functions of EEG device such as signals digitizing and preprocessing, buffering, calibration, communication with PC for configuration and data transmitting, etc. are imposed on a so-called firmware, which is the microcontroller software program. Modern microcontrollers are sufficiently universal and usually have built-in ADC and/or signal processing commands. Typical implementations of EEG-device interact with a personal computer, where data goes through analytical, visualization software, and communicates with other applications. In this case it should support one or several interfaces, such as USB or Bluetooth.

Digitized data can be transmitted to the computer with certain frequency, limited by the interface speed. Electroencephalography uses various sampling rates. Most commonly used rate is near 250 Hz, but in some applications it comes up to several kHz and more. The depth of information we can obtain by analyzing EEG channels depends on a sampling rate. With insufficient frequency a result of analysis will be distorted which decreases the system's efficiency.

## 2. An Analysis of Brain Signals

According to the guidance of many works, Brain-Computer Interface is an approach to computer control, a new kind of Human-Computer Interface. It analyses signals, produced by the brain (bioelectric potentials) during various cognitive, emotional and physiological processes, to make automated intentional influence on the behavior of computer systems and software applications (generation of control commands). Various hardware types can be used as a source of data for further BCI analysis, such as MRI or infrared laser scanners, but the most attractive with a view to the responsiveness of an interface is an EEG apparatus. The latter is the subject of present paper.

A BCI-system is, in general, composed of the following components: *signal acquisition, preprocessing, feature extraction, classification (detection), post processing and application interface.* The signal acquisition component is responsible for recording the electrophysiological signals and providing the input to the BCI. On-line BCI systems use EEG to measure brain potentials (instead of MRI, for example) because of its flexibility. Preprocessing includes artifact reduction (electrooculogram (EOG) and electromyogram (EMG)), application of signal processing methods, i.e. low-pass and / or high-pass filter, methods to remove the influence of the line-frequency and in the use of spatial filters (bipolar, Laplacian, common average reference). After preprocessing, the signal is subjected to the feature extraction algorithm. The goal of this component is to find a suitable representation (signal features) of the electrophysiological data that simplifies the subsequent classification or detection of specific brain patterns. There are a variety of feature extraction methods used in BCI systems; a non exhaustive list of these methods includes amplitude measures, band power, Hjorth parameters, autoregressive parameters and wavelets. The task of the classifier component is to use the signal features provided by the feature extractor to assign the recorded samples of the signal to a category of brain patterns. In the simplest form, detection of a single brain pattern is sufficient. This can be achieved by using a threshold method. More sophisticated classifications of different patterns depend on linear or nonlinear classifiers. Post-processing issues such as dwell time (time in a certain state before an event occurs), refractory period (time after an event), combination of classifier and time dependent modeling uses the pre-knowledge of the actual experiment to adapt the classifier output to working conditions (specific person, environment, etc.). In the case of an asynchronous or self-paced BCI the non-control state has to be identified alongside with command states. Hereupon a complexity of analysis increases, requiring it to be more sensitive to the electrical indications' changes. Not only very command should be determined, but also user's intention to execute a command and intensity of one. The final output of the BCI is transformed into an appropriate signal that can then be used to control a Virtual Environment (VE) system, prosthetic or other kind of application.

It should be noted also that EEG signals are composed of different oscillations named "rhythms" (e.g. alpha, beta, gamma, delta, theta, and mu) which have distinct properties in terms of spatial and spectral localization. Particular frequency bands display individual relation to the emergence of percepts, memories, emotions, thoughts, and actions. Together with distribution of functions across the areas of the cortex it gives BCI a means to concentrate on a specific type of neural activity of a brain.

## 3. An Adaptability of BCI

Saying about BCI, it is good to start with influences which result in electrical brain responses. Neurofeedback is a fundamental element of BCI. Theory defines the term Evoked Potentials (EP) which implies a recording of electrical oscillations of nervous system of a human or animal following the presentation of external stimulus. Event-Related Potential (ERP) is more neutral term meaning bioelectrical responses related not only to evoked signals but also to invoked i.e. appeared after an internal stimulus [2]. The latter can be caused for example by higher nervous activity. These responses are time-locked to the irritant and can tell us about the behavior of mental processes.

Movement preparation followed by execution, or even only motor imagination is usually accompanied by a power decrease in certain frequency bands, labeled as event-related desynchronization (ERD), in contrast, their increase after a movement indicates relaxation and is due to synchronization in firing rates of large populations of cortical neurons (ERS). Such indicators are very useful in real-time EEG analysis.

*Synchronous BCI* should provide the user an ability to interact with the targeted application only during specific time periods, imposed by the system. Hence, the system informs the user, thanks to dedicated stimuli (generally visual or auditory), about the time location of these periods during which they has to interact with the application. The user has to perform mental tasks during these periods only. If he or she performs mental tasks outside of these periods, nothing will happen.

Instead of this, in *asynchronous (self-placed) BCI* the user can produce a mental task in order to interact with the application at any time. He or she can also choose not to interact with the system, by not performing any of the mental states used for control. In such a case, the application would not react (if the BCI works properly). Naturally, self-paced BCI are the most flexible and comfortable to use. Ideally, all BCI should be self-paced. However, it should be noted that designing a self-paced BCI is much more difficult than designing a synchronous BCI. With a self-paced BCI, the system has to analyze continuously the input brain signals in order to determine whether the user is trying to interact with the system by performing a mental task. If it is the case, the system has also to determine what mental task the user is performing. Designing an efficient self-paced BCI is presently one of the biggest challenges of the BCI community and a growing number of groups are addressing this topic.

*Operant conditioning* is a technique according to which a user during relatively long period of time learns how to use BCI. In that time, interface adjusts its algorithms and parameters to particular user, increasing speed and quality of analyzing procedure. A subject learns how to modify voluntarily characteristics of their EEG. Meanwhile, psychology knows that long training usually results in development of conditioned reflex. One of major courses of current BCI research is to develop an *unconscious operant conditioning* which is sensitive to a phenomenon of psychological automatism. In the paper [6] we can read about the experiment with automotive color selection by means of user's simultaneous voluntary change of characteristics of 3 EEG rhythms (for RGB model). A prerequisite to the experiment was the fact that surrounding color can modify mental state in definite direction on subconscious level. A subject connected to a BCI looks at the computer screen which displays certain background color. After some time (trial period) a color on the screen changes to the most preferable for this person. As we see, color here plays a role of feedback. It's changing helps brain to discover the right way to influence on the BCI program. And whole process occurs at the level of intuitive intentions.

Various authors have described examples of unconscious conditioning and cognitive feedback. There are many studies of modified mind states, hypnosis, meditation and their echo on the different parameters of EEG. Most of such components are in stable proportion to the time scale, for example, P300, MERMER components, etc. These are indicators convenient for detection of a type of a progressing mental activity. Segment EEG characteristics (mutual reaction of ensembles of neurons on certain stimulus) is a conceptual point in describing of integrative and systematic properties of brain dynamics. Signal patterns open a way to most subtle EEG analysis. But this apparatus is not so unambiguous.

It is important to provide a high degree of adaptability, which should reflect in quick learning (operant conditioning). BCI should be sufficiently reliable to determine real user's intentions. It should successfully work with all manner of people and must not depend on emotional conditions, artifacts and surrounding influences. Naturally, we can find an application for fuzzy logic here.

## 4. Electronic learning system

SignSpace is an author's project directed to research a possibility of BCI application in education process. The principal goal is to investigate a process of information understanding. Main part of this learning system is client software which should contain a realization of BCI interface. This software must support various types of computer devices (PC, mobile phones, notebooks, communicators, etc.). As an educational system it should have controls to represent learning materials. Software must include communication capabilities to interact with server part which is described further. The module mostly related to the topic of this paper is a neurofeedback component.

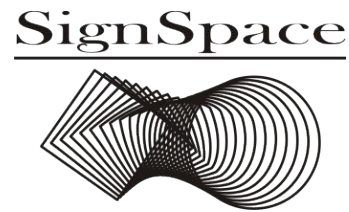

Server part works as a control core of a system and can provide integration of many clients into a corporative learning media. Several functions of it are to store and manage all educational data and collect formal and organizational information concerning learning processes (curriculum, schedules, registration data of users, trees of dependencies etc.), keeping analytical data obtained from clients (EEG) during an interaction with users, to perform off-line processing of this data with further utilization of valuable information obtained. Server should have output to the distributed computing environment such as GRID to provide regularities research in the raw EEG data to improve algorithms of BCI analysis. In general, server part should play a role of research center of whole system giving continuous improvements and development to it.

Studying process represent the next procedure. A subject takes their place, connects to the BCI interface and starts to play a video lecture. BCI using a special programme monitors a brain activity of the user during cognition of video material. System monitors a degree of clearness of obtaining information. When system determines student's incomprehension it stops a video lecture and shows them additional advisory information or changes a way of information display. These operations have a purpose to make information completely clear to the user who studies it.

Of course it is necessary to use specific methods of lectures design, evaluation of a degree of learning successfulness, taking practical lessons. But great advantage of proposed system in comparison with other types of e-learning are high adaptability to the user and possibility of individual automatic approach to information presentation.

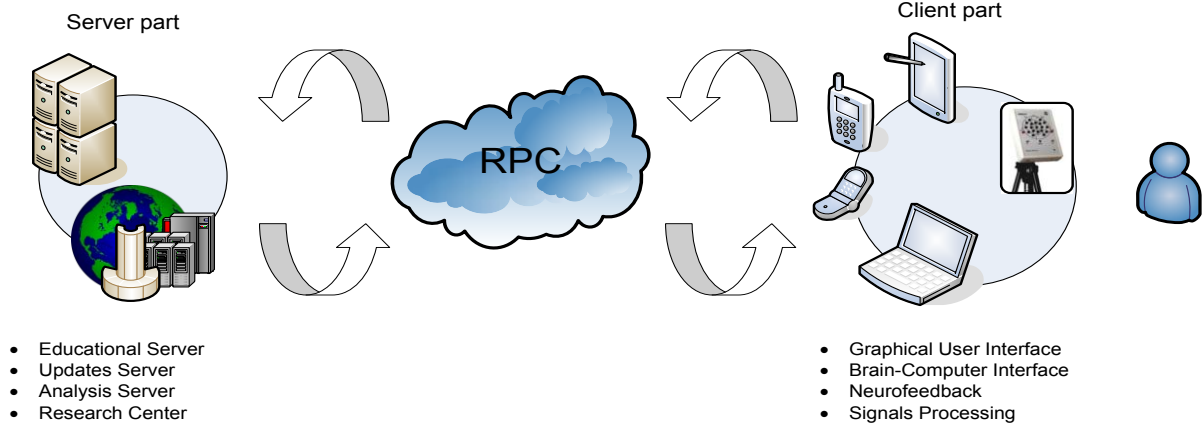

- Educational Server
- Updates Server
- Analysis Server
- Research Center

- Graphical User Interface
- Brain-Computer Interface
- Neurofeedback
- Signals Processing

## 5. The process of cognition

Consider mind as a display of profound laws of nature. In any field of action we can meet such problems as a lack of information, complexity and correlations. And everywhere we can study optimality by observing various kinds of interactions. No matter what kind of space and forces we choose to investigate, the phenomenon of interactions is always inherently the same. This is a part of more fundamental problem.

Making a chain of logical inferences using various notions is usually characterized as enclosure and inheritance of concepts. But we cannot construct an exhaustive description of investigated system by merely moving from one point to another. That's why the main question remains unresolved: what is the phenomenon of interaction, what its nature and place among objective laws. So-called "laws of formal logic" are able to describe only particular relations between objects, not a general view. The nature of things is closely connected with spatial relations between objects where each of them plays its own role. We should find missed parts in current description of logical law. Space limitations [8] and making a choice are the key points of this challenge.